\documentclass[showpacs,preprintnumbers,amsmath,prl,
graphicx,amssymb,superscriptaddress,twocolumn,nofootinbib]{revtex4}
\usepackage{graphicx}
\usepackage{dcolumn}
\usepackage{bm}

\renewcommand{\today}{\number\day\space\ifcase\month\or January\or 
 February\or March\or April\or May\or June\or July\or August\or 
 September\or October\or November\or December\fi\space\number\year}
 
\renewcommand{\labelitemi}{$\bullet$}

\begin{document}
\title{A Realistic Assessment of the Sensitivity of XENON10 and 
XENON100 \\ to Light-Mass WIMPs}

\newcommand{\efi}{Enrico Fermi Institute, Kavli Institute for 
Cosmological Physics and Department of Physics,
University of Chicago, Chicago, IL 60637}
%%%%%%%%%%%%

\affiliation{\efi}
														
\author{J.I. Collar}\affiliation{\efi}

\begin{abstract}

The underlaying assumptions and uncertainties involved in the derivation of 
WIMP exclusion limits from XENON10 and XENON100 detectors are 
examined. In view of these, recent 
claims of sensitivity to light-mass Weakly Interacting 
Massive Particles (WIMPs) are shown to be overstated. Specifically, bounds constraining regions of interest in WIMP 
parameter space from the DAMA/LIBRA, CoGeNT and CRESST experiments can be assigned only 
a very limited meaning. 

\end{abstract}
%% ENTER PACS NEXT
%\pacs{85.30.-z, 95.35.+d, 95.55.Vj, 14.80.Mz}

\maketitle

%%%%%%%%%%%%%%%%%%%%%%%%%%%%%%%%%%%%%%%%%%%%%%%%%%%%%%%%%%%%%%%%%%%%%%

The possibility of relatively-light ($<$10 GeV/c$^{2}$) Weakly 
Interacting Massive particles (WIMPs)
being the source of the  annual modulation effect 
\cite{andrzej} observed in the DAMA/LIBRA experiment \cite{dama} was first 
proposed in \cite{pglight} and later revisited in \cite{kzlight}. 
Since then, it has gained in popularity 
as constraints from other searches have depleted alternative dark matter 
scenarios, and several particle phenomenologies have generated 
plausible candidates in this mass range. The situation has recently gained in 
complexity with the observations from CoGeNT 
\cite{ourprl2,modcogent} and CRESST \cite{cresst}, which 
may point at a light-WIMP parameter space compatible with DAMA/LIBRA. 

Studies of detector sensitivity to such light dark matter particles 
must be regarded with caution \cite{later}. For presently-available technologies, 
light-WIMP signals would fall uncomfortably close to detector thresholds, a 
region where systematic effects can lead to rushed claims of 
exclusion or detection. Experimentalists should attempt 
not to aggravate what is a naturally murky area of study, by 
describing the assumptions made to arrive to their 
conclusions, and by including an assessment of the uncertainties 
involved in their analyses. Recent work by the XENON10 \cite{xenon10} and XENON100 
\cite{xenon100,plante} collaborations is lacking in these 
respects. The goal of this report is to provide this missing 
information as succinctly as possible. 

\section{I: Recent measurement of $\mathcal{L}_{\text{eff}}$ by G. 
Plante {\it et al.} \cite{plante}.}

A previous attempt to derive light-WIMP limits by the XENON100 
collaboration \cite{xenon100p} was received with criticism
\cite{juandan} pointing out the unphysical behavior of the quenching 
factor in the production of direct scintillation by nuclear recoils 
($\mathcal{L}_{\text{eff}}$) employed. An additional critique 
\cite{later} emphasized traceable mistakes made in the data analysis 
of previous $\mathcal{L}_{\text{eff}}$ measurements performed 
by the XENON10 and XENON100 collaborations. 

A new recent $\mathcal{L}_{\text{eff}}$ measurement 
by the XENON100 collaboration (Plante {\it et al.} \cite{plante})
addresses the concerns in \cite{later}, specifically the systematic 
effect introduced by erroneously normalizing simulated recoil rates 
to their corresponding
measured values, and the sub-optimal design of several earlier 
calibration detectors, prone to 
multiple scattering involving inert materials. The new detector used in \cite{plante}
features a compact design that bypasses this concern and maximizes
light collection from the active liquid xenon (LXe) volume. 
Not surprisingly, the monotonically decreasing behavior of 
$\mathcal{L}_{\text{eff}}$
towards zero energy predicted in \cite{juandan,later} is now observed 
by Plante {\it et al.}

While great strides towards a better understanding of 
$\mathcal{L}_{\text{eff}}$ have been made in \cite{plante}, significant room for 
improvement remains:

\begin{list}{\labelitemi}{\leftmargin=0.em}
    
\item An unnecessary degree of freedom in the fits comparing 
LXe scintillation measurements and 
simulations has been introduced by Plante {\it et al.}, 
namely the energy resolution as a function of recoil 
energy, which is a predictable quantity, and not independent of $\mathcal{L}_{\text{eff}}$, as implicitly postulated in \cite{plante}. 
This is in 
contrast to an earlier measurement by Manzur {\it et al.} 
\cite{manzur}, also correctly pointing at a decreasing 
$\mathcal{L}_{\text{eff}}$, where the 
resolution was determined by measurements at energies well-above any threshold 
effects, and for all lower energies unambiguously 
defined according to its expected dependence on photoelectron yield (a function of $\mathcal{L}_{\text{eff}}$). 
The introduction of this gratuitous degree of freedom can 
reinstate the 
deleterious effects described in \cite{later}, by substantially 
biasing $\mathcal{L}_{\text{eff}}$ towards artificially large values and reducing uncertainty. 
This concern is particularly important below 
$\sim$6.5 keV$_{r}$, where threshold effects become dominant in \cite{plante}. The approach 
taken in \cite{plante} does not necessarily
have to constitute an issue, as long as the obtained best-fit
resolution follows the expected 
behavior$^{1}$\footnotetext[1]{Unfortunately, a sizable mismatch 
seems to be involved \cite{plantepriv}.}. No mention of this comparison is made 
in \cite{plante}. As described in Sec.\ II below, the 
extrapolated behavior of $\mathcal{L}_{\text{eff}}$ to zero energy 
critically 
determines LXe sensitivity to light-WIMPs, making attention to such details 
very important. A discussion of this comparison between expected 
and best-fit energy resolution would considerably improve 
the credibility of the lowest-energy $\mathcal{L}_{\text{eff}}$ 
values obtained by Plante {\it et al.}.

\item Measurements in \cite{plante} were performed in single-phase 
mode, i.e., in the absence of the electric drift field present during 
the operation of the XENON100 detector. This field is expected to 
suppress electron-ion recombination, reducing the scintillation yield. 
While this effect was found to be small by Manzur {\it et al.} 
\cite{manzur}, the $\mathcal{L}_{\text{eff}}$ values by Plante {\it 
et al.} should 
be considered an upper limit to the actual nuclear recoil 
scintillation yield in the 
XENON100 detector. This consideration as an upper limit is revisited 
in Sec.\ II within a different  
(instrumental) context. 

\item It must be kept in mind that the definition of 
$\mathcal{L}_{\text{eff}}$ used by the LXe detector 
community differs from the traditional one for a quenching factor, 
by relativizing the scintillation yield from 
low-energy nuclear recoils to that from electron recoils at a relatively high 
ionization energy (122 keV). The more conventional definition uses
the ratio of scintillation yield from nuclear and 
electron recoils of identical energy. This may seem like a moot point, until 
the large non-proportionality typically observed in heavy 
scintillators \cite{scint}, including LXe \cite{scint2}, is 
examined: a large increase in scintillation yield for electron 
recoils (the denominator in the traditional definition of quenching factor) is typically 
observed below few hundred of keV down to few keV. Compton scattering measurements complementary to those in \cite{plante} are clearly advisable.

\end{list}

\section{II: New light-WIMP limits from XENON100 \cite{xenon100}}

The analysis of a 100 day exposure from the XENON100 detector 
\cite{xenon100} has resulted in a claim of sizeable improvement in light-WIMP 
sensitivity with respect to a previous shorter (11 day) run \cite{xenon100p}.
A discussion of the strong  assumptions implicitly made to arrive to 
this conclusion and of the 
uncertainties neglected in 
\cite{xenon100} is provided below.

\begin{figure}
\includegraphics[width=8.5cm]{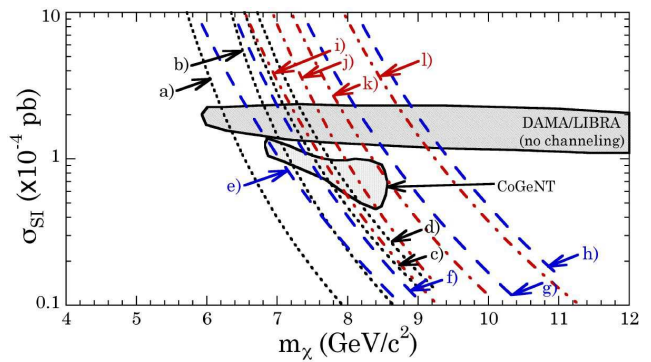}
\caption{90\% C.L. XENON100 exclusion contours obtained under the assumptions 
discussed in the text. Dotted (black) curves correspond to the 
$\mathcal{L}_{\text{eff}}$ by Plante {\it et al.} \protect\cite{plante}, dashed (blue) to that by Manzur {\it et al.} 
\protect\cite{manzur}, 
and dash-dotted (red) to a most recent Monte Carlo-independent $\mathcal{L}_{\text{eff}}$ determination by ZEPLIN-III \protect\cite{zeplin3}. The notation used to describe each 
case lists number of irreducible recoil events accepted, $\mathcal{L}_{\text{eff}}$ 
considered (central value of logarithmic extrapolation or its lower C.L. boundary), and statistics used (see 
text): a) 1 event, central, Poisson; b) 1 ev., 2$\sigma$, Poisson; c) 4 evs., 2$\sigma$, Poisson; d) 4 evs., 2$\sigma$, binomial; e) 1 
ev., central, Poisson; f) 1 ev., central, binomial;  g) 1 ev., 
1$\sigma$, Poisson; h) 4 evs., 1$\sigma$, 
binomial; i) 1 ev., central, 
Poisson; j) 1 ev., central, 
binomial; k) 4 evs., central, 
binomial; l) 4 evs., 1$\sigma$, 
binomial. Additional instrumental uncertainties not 
reflected in this figure are listed in the text. 
}
\end{figure}

\begin{list}{\labelitemi}{\leftmargin=0.em}
    
\item Contour ``a'' in Fig.\ 1 is similar to the exclusion curve in 
\cite{xenon100}. It is obtained by assuming the 
logarithmic extrapolation of the $\mathcal{L}_{\text{eff}}$ from 
Plante {\it et al.}, as proposed in 
\cite{xenon100}, and that only 
the lowest in energy
of the three accepted nuclear-recoil events in \cite{xenon100} could 
be due to a light-WIMP. Contour ``b'' in the same curve represents the 
exclusion obtained when the 2$\sigma$ C.L. uncertainty band in this 
$\mathcal{L}_{\text{eff}}$ is adopted instead. The resulting change 
in sensitivity is much larger than what is indicated by the 
very narrow uncertainty bands in Fig.\ 5 of \cite{xenon100}. This 
issue can be confirmed by performing a self-consistency test between 
the XENON100 exclusion curves in \cite{xenon100} and \cite{xenon100p}: the two 
values of $\mathcal{L}_{\text{eff}}$ contemplated in \cite{xenon100p} 
(Fig.\ 1 there) 
generated exclusion curves diverging by a very large factor for light WIMP 
masses (Fig.\ 5 in \cite{xenon100p}). Those two values of 
$\mathcal{L}_{\text{eff}}$ are coincidentally not very different from 
the central and 2$\sigma$ C.L. boundaries of the $\mathcal{L}_{\text{eff}}$ from 
Plante {\it et al.} (Fig.\ 1 in \cite{xenon100}). However, in Fig.\ 5 
of \cite{xenon100}, the new 
XENON100 
analysis assigns an insignificant impact on the exclusions to this large spread in 
$\mathcal{L}_{\text{eff}}$. 
The origin for this lack of self-consistency must be 
addressed$^{2}$\footnotetext[2]{This question may be extended to 
higher WIMP masses. For those, following basic statistical estimators, 
only a marginal increase in 90\% C.L. sensitivity 
should be expected in going from zero to three irreducible events 
following an increase in exposure by a factor of ten. A much larger 
gain in heavy WIMP sensitivity has been claimed in going from \cite{xenon100p} to \cite{xenon100}.}. 

\item In a departure from the blind analysis initially intended by the 
XENON100 collaboration, 
three events next to threshold were rejected immediately following unblinding 
\cite{xenon100,nyt}. These events have been ascribed to photomultiplier (PMT) noise affecting 
the S1 (direct scintillation) channel. Post-unblinding corrective actions are 
often required and no judgment on this decision 
should be passed until more details become available. However, it is 
worth remembering that this type of PMT noise was already present in XENON10 data 
\cite{10noise} (and not rejected a posteriori) 
and is also ubiquitous in a XENON100 example event catalogued as 
``good'' \cite{laura}, indicating that data cuts originally deemed as 
adequate must have been in place against 
it$^{3}$\footnotetext[3]{A single example of noise-correlation provided 
in \cite{laura} corresponds to an event 
well-below threshold (3.1 S1 photoelectrons) and does not seem to display the 
S1 region corresponding to the S2 pulse.}. Details about the post-unblinding 
criteria developed to reject these events while in the presence of robust S2 
(ionization) signals will be of special interest. Contours labelled ``c'', 
``d'' and ``h'' in Fig.\ 1 display the non-negligible effect of including these 
three rejected events into the calculation of XENON100 exclusions.

\item Dashed lines in Fig.\ 1 represent XENON100 exclusions using a 
logarithmic extrapolation of the 
alternative $\mathcal{L}_{\text{eff}}$ obtained by Manzur {\it et 
al.} \cite{manzur} using another optimally-designed LXe chamber. 
Differences in data treatment and mode of 
operation between \cite{plante} and \cite{manzur} are 
described above. Criticisms concerning the data treatment in 
\cite{manzur} are given in \cite{later}. Contours labelled ``g'' 
and ``h'' use the logarithmic extrapolation of the lower 1$\sigma$ 
C.L. boundary rather than the central  $\mathcal{L}_{\text{eff}}$ 
value (contours ``e'' and ``f'').

\item A lingering critical question is to what extent a determination of 
$\mathcal{L}_{\text{eff}}$ performed using highly-optimized compact 
calibration detectors like those in \cite{plante,manzur} can be 
applied with confidence to a much larger device like the XENON100 detector, featuring 
a small S1 light-detection efficiency (just $\sim$6\% 
\cite{raf}), different hardware trigger 
configuration, data processing, etc. 
For instance, simulations like those used 
within XENON100 to obtain a cumulative cut acceptance near threshold can only be 
regarded as best-effort estimates. Their limited meaning and tendency 
to significantly overestimate near-threshold efficiency has been 
recently encountered by Plante {\it et al.} \cite{plante}, even for the near-ideal 
conditions of their small calibration chamber ($\sim$18 c.c. of LXe, 
with 4$\pi$ PMT coverage). Another example of 
instrumental constraints is 
the negligibly small low-energy {\it effective} $\mathcal{L}_{\text{eff}}$ 
derived by the 12 kg ZEPLIN LXe 
dark matter detector \cite{zeplin} when its measurement 
is attempted {\it in situ} (if adopted, this $\mathcal{L}_{\text{eff}}$ 
generates essentially no LXe sensitivity to WIMP masses below 10 GeV/c$^{2}$ 
\cite{juandan}). Going back some time, a dramatic deficit in 
observed neutron-induced recoil rates compared to few-keV$_{r}$
expectations was observed with the XENON10 detector \cite{manzuraps}. 
Such comparisons should be revisited within the context of the 
existing XENON100 neutron calibrations: 
if the expected response to few-keV$_{r}$ recoils is still absent due to instrumental limitations, 
light-WIMP limits should not be distilled from sheer wishful 
thinking. As discussed next, light-WIMP limits are obtained by XENON100 
under the strong assumption of positive Poissonian fluctuations in 
scintillation light from WIMP-induced recoils well-below the detector 
energy threshold. The agreement between expected and observed neutron-induced 
recoil rates should therefore 
be demonstrated into that energy range. The XENON100 
collaboration is invited to produce this much needed validation of 
their claims$^{4}$\footnotetext[4]{This request for validation is not 
unfair: COUPP results consistently include a sensitivity penalty 
whenever any disagreement with neutron-induced recoil rate
expectations arises \cite{coupp}. A CoGeNT detector 
presently installed at Soudan has not been exposed 
to neutron sources to avoid activation during a search for an annual 
modulation \cite{modcogent}. However, the response to sub-keV$_{r}$ nuclear 
recoils was
measured with a detector identical in properties (energy resolution, noise, 
bias) and crystal mass 
\cite{cogentcalib}.}.

\item The mentioned assumption of Poissonian statistics 
governing the 
microscopic processes$^{5}$\footnotetext[5]{Notice reference is made 
to those {\it ab initio} processes and not the ensuing statistics of 
photoelectron generation in PMTs.} of light and charge 
generation by few keV$_{r}$ nuclear 
recoils in LXe is not only presently unwarranted \cite{juandan}, 
but seemingly counter to the scarce available 
experimental information: the very small value of the Fano factor in LXe 
is for instance indicative of sub-Poissonian statistics ruling those 
processes \cite{fano}. Similarly, 
the electron emission statistics by few-keV, heavy-mass 
ions during surface collisions is known to be better described by binomial 
rather than Poissonian
statistics \cite{binomial}. Contours ``d'', ``f'' and ``h'' in Fig.\ 1 
display the effect of a small deviation from the Poisson 
assumption (a binomial distribution of same mean, taking a probability of 
S1 photon detection of 6\% \cite{raf}). These contours should be considered  
as  illustrative ansatzes for information-carrier statistics that could generate even less 
light production. In this respect, it is worth emphasizing that a 7 GeV/c$^{2}$ 
light-WIMP is expected to impart a mean recoil energy in LXe of just 
$\sim$0.6 keV$_{r}$, with an absolute maximum (occurring with infinitesimal
probability) of $\sim$4.1 keV$_{r}$. At this very endpoint, the 
probability of surpassing the XENON100 four-photoelectron ($\sim$8.5 
keV$_{r}$) threshold is 
never larger than $\sim$10\%, even when the logarithmic extrapolation 
of the
$\mathcal{L}_{\text{eff}}$ by Plante {\it et al.} 
and Poisson fluctuations are adopted. 
Not all light-WIMP detecting media fare as poorly
from this point of view of generation of information carriers: the 
same 7 GeV/c$^{2}$ WIMP at its spectral endpoint ($\sim$1.4 keV 
ionization) in CoGeNT germanium diodes would generate a readily 
detectable $\sim$470 electron-hole pairs. 

\item Fig.\ 1 includes the present uncertainty in the quenching factor 
for sodium recoils in DAMA/LIBRA \cite{juandan,danq}, a subject of 
discussion avoided by the 
XENON100 collaboration in \cite{xenon100,xenon100p}. This uncertainty 
extends the DAMA/LIBRA region to considerably lower WIMP masses than what 
is represented in \cite{xenon100,xenon100p}. 

\end{list}

\section{III: Light-WIMP limits from XENON10 via ionization signals 
\cite{xenon10}}

A recent reanalysis of XENON10 data uses strictly the 
ionization channel in that detector to impose limits on light-mass 
WIMPs \cite{xenon10}. This S2 light emitted via electroluminescence 
from charge drifted into the gas phase of the device is, when examined alone, 
sufficient to extend the sensitivity of LXe detectors to recoils of 
O(1) keV$_{r}$. Not including the information from S1 (direct 
scintillation) allows a reduction in threshold at the expense of 
losing the ability to distinguish nuclear from electron recoils.

While this approach is promising, the few keV$_{r}$ nuclear recoil energy 
scale corresponding to this S2 channel is 
presently hopelessly ill-defined. This is a result of the  inadequate 
``best-fit Monte Carlo''  method \cite{sorth} employed to arrive at 
it. An extensive critique of this method, ignored thus far by XENON10 authors, can be found in 
\cite{later}. In 
a troubling case of double-standards, the gist of this 
critique (that with this method all uncertainties are absorbed into the 
energy scale) has been recently echoed by the XENON100 collaboration (Sec. I in \cite{plante}), 
when rebuking 
indirect measurements of $\mathcal{L}_{\text{eff}}$ using the same 
methodology. 

In lieu of a reiteration of the criticisms in \cite{later}, the reader is invited to inspect 
Fig.\ 2: each of the colored energy scales shown there, generated by 
the ``best-fit Monte Carlo'' method, has been claimed to be the 
correct one by the XENON10 collaboration over the brief span of the last
two years. The scale is observed to change as rapidly as from 
workshop presentation to its published proceedings. Its monotonic evolution has 
been towards the black curves, held in \cite{later} to correspond 
to the most plausible energy scale (one derived from an earlier method laid out by XENON10 authors: see pertinent discussion in \cite{later}). A critically-minded reader would (rightly) argue that none of these can be presently assigned any 
credibility at few keV$_{r}$. However, reasons have been provided in \cite{later}
to support the solid black curve, representing the Lindhard theory 
modified below a $\sim$40 keV$_{r}$
kinematic threshold \cite{juandan} by an 
example of adiabatic correction, as in \cite{ahlen}:
 
\begin{list}{\labelitemi}{\leftmargin=0.em}
    
\item This energy scale generates a similar quenching for the 
ionization yield and the $\mathcal{L}_{\text{eff}}$ observed by 
Plante {\it et al.} or Manzur {\it et al.}, i.e., a monotonic 
decrease in the generation of information carriers (free charge, 
direct scintillation 
photons) below kinematic 
threshold \cite{juandan}, having an effective cutoff at $\sim$1 keV$_{r}$. 
That both processes should decrease hand-in-hand can be argued 
based on the dominant role of ionization as the main precursor to direct 
scintillation for low energy recoils in LXe \cite{juandan,doke}.

\item The energy scales postulated thus far by XENON10 (color 
curves in Fig.\ 2) overestimate, by several orders of magnitude, 
the very small average charge yields observed in impact ionization 
experiments involving few keV and sub-keV xenon ions. The relevance of these 
measurements and examples from the literature are discussed in \cite{later}, 
where it is also emphasized that XENON10 workers are not without a 
reference on what to expect at these low energies. The introduction 
of the adiabatic 
correction proposed in \cite{ahlen} resolves this disagreement. 

\end{list}

\begin{figure}
\includegraphics[width=7.5cm]{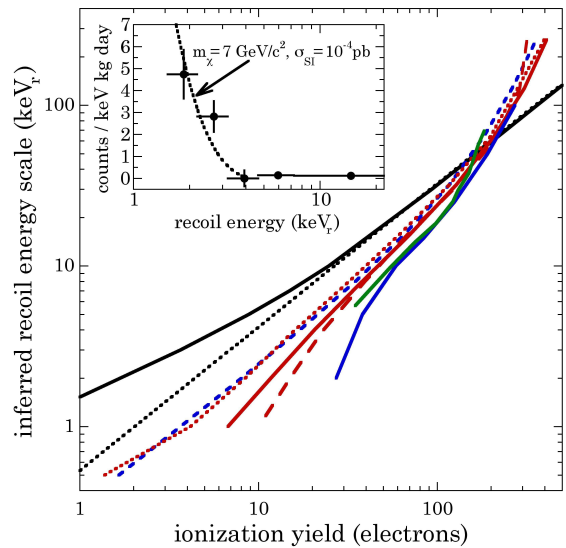}
\caption{Updated and adapted from \protect\cite{later}. Color
lines correspond to the rapid evolution of the XENON10 S2 ionization energy scale over 
the past two years, using the ``best-fit Monte Carlo'' method 
\protect\cite{sorth}. In 
chronological order: solid blue and green \protect\cite{sorth}; 
dashed red \protect\cite{petertaup1} (workshop presentation);
solid red \protect\cite{petertaup2} (its proceedings); 
dotted red, 1$\sigma$ C.L. contour in \protect\cite{petertaup2}, as per 
\protect\cite{xenon10}; 
dashed blue \protect\cite{xenon10,petereric}.
%Datapoints from \protect\cite{manzur} appearing in the original for 
%this figure
%and claimed in \protect\cite{later} to be due to threshold effects
%have been acknowledged as such in \protect\cite{xenon10,peterjpac}.
Dark lines correspond to the predicted energy scales in 
\protect\cite{later} (see text). Inset: alternative analysis of the 
few-electron S2 background in XENON10 \protect\cite{xenon10} (see text).
}
\end{figure}

Of special interest is a population of XENON10 low energy ionization signals 
described in \cite{xenon10} as single 
electrons$^{7}$\footnotetext[7]{The well-know  
ionization ``afterpulses'' following large energy depositions in LXe, limited to 
single electrons but nevertheless mentioned in 
\cite{xenon10} as a possible origin for these multi-electron events, can be trivially removed 
through a $\sim$100$\mu$s delayed-coincidence cut 
\cite{later,raf,zepspon}. No mention is made of the application of this cut 
to the dataset in \cite{xenon10}. This should be 
clarified.}. This definition is 
both surprising and misleading. As mentioned in \cite{later}, the 
large amplification gain provided by the electroluminescence provides 
a good resolution in the multiplicity of drifted charge in LXe. These events 
clearly include a multi-electron component and their population has been described as 
such by the contact author of \cite{xenon10}, as recently as in 
\cite{petertaup2}. Their origin is unknown, and hard to ascribe to 
minimum ionizing particles in an efficient self-shielding medium such as 
LXe \cite{later}. While none of this is discussed in \cite{xenon10}, 
their accumulation towards low values of S2 pulse-width 
is to be expected from the effect of charge multiplicity on this 
variable, and does not have to correspond to a radioactive 
contamination close to the $z\!=\!0$ detector coordinate (this is unlikely, given 
that roughly the same number of PMTs, major sources of internal 
activity in the XENON10 detector, are placed at both extrema of $z$). 

\section{IV: Conclusions}

The inset in Fig.\ 2 represents the differential rate of few-electron 
S2 events in XENON10, obtained by applying the same five background 
cuts as in \cite{xenon10}, extending the analysis down to the S2 = 1e$^{-}$ boundary, and adopting the energy scale described by the 
black solid line in the same figure. This differential rate seems 
to be also compatible with Fig.\ 2 in \cite{petertaup2}, once the 
adopted energy scale is included. It offers a good match to the 
expected signal from a light WIMP in the region of interest of other 
searches (DAMA/LIBRA, CoGeNT, CRESST). The outcome of this exercise, 
performed here strictly for the sake of argument, 
should be strongly de-emphasized at this time, given 
the present lack of knowledge about this energy scale evidenced in Fig.\ 2. 
The reader should remember instead that a further evolution by 
a mere $\sim$1 keV$_{r}$ in 
the ever-changing S2 energy scales postulated by XENON10 can transform 
the ``severe constraints"  of  \cite{xenon10} into a signal in principle 
compatible with a light-WIMP. 

In conclusion, the claims in \cite{xenon10} are clearly presently 
untenable. Awaiting clarification of the several pending issues 
pointed out in Sec.\ I and II, light-WIMP limits obtained through a more conventional 
analysis of XENON100 data \cite{xenon100} can only be assigned the very limited meaning 
illustrated by Fig.\ 1. The XENON100 collaboration is congratulated for 
the recent advancement in their understanding of 
$\mathcal{L}_{\text{eff}}$, encouraged to develop improved methods of 
characterization of the S2 energy scale leading to a reliable 
exploration of light-WIMP candidates, 
and urged to employ transparency in the discussion of 
uncertainties and assumptions underlaying their results, in view of 
the very limited performance of LXe as a light-WIMP detection medium. 
Finally, while several interesting phenomenological routes to 
alleviate tension between LXe constraints and other light-WIMP 
searches have been put forward recently \cite{alev}, these deviations 
from arguably more conventional assumptions do not seem to be mandatory at this time.

N.B.: A new measurement of $\mathcal{L}_{\text{eff}}$ by the ZEPLIN-III collaboration appeared coincident with the release of this preprint \cite{zeplin3}. Fig.\ 1 now reflects XENON100 exclusions obtained with it. ZEPLIN-III derives a $\mathcal{L}_{\text{eff}}$ decreasing below 40 keV$_{r}$ and vanishing at few keV$_{r}$, in tight agreement with the LXe kinematic threshold described in \cite{juandan}.

\end{document}